\documentclass{aa}
\usepackage{graphicx}
\usepackage{amssymb}

\begin{document}
   \title{Dynamical Mass Estimates for Two Luminous Star Clusters
   in Galactic Merger Remnants \thanks{Based on Observations at the
   Very Large Telescope of the 
   European Southern Observatory, Paranal/Chile under Program
   073.D-0305(B). }
}  

   \titlerunning{Dynamical masses of star clusters in mergers}

   \author{N. Bastian$^{1,2}$, R.P. Saglia$^{3}$, P. Goudfrooij$^{4}$, M. Kissler-Patig$^{2}$,C. Maraston$^{5}$, F. Schweizer$^{6}$, and M. Zoccali$^{7}$ 
          }

\authorrunning{Bastian et al.}

   \offprints{bastian@star.ucl.ac.uk}

   \institute{$^1$ Department of Physics and Astronomy, University College London, Gower Street, London, WC1E 6BT \\
	      $^2$ European Southern Observatory, Karl-Schwarzschild-Strasse 2 
               D-85748 Garching b. M\"{u}nchen, Germany \\              
	       $^3$ Max-Planck-Institut f\"{u}r Extraterrestrische
	      Physik, Giessenbachstrasse, D-85748 Garching, Germany \\
	      $^4$ Space Telescope Science Institute, 3700 San Martin
	      Drive, Baltimore, MD 21218, USA \\
	      $^5$ University of Oxford, Denys Wilkinson Building,
              Keble Road, Oxford, OX13RH, United Kingdom \\
	      $^6$ Carnegie Observatories, 813 Santa Barbara Str.,
	      Pasadena, CA 91101-1292, USA \\
	      $^7$ Pontificia Universidad Cat\'{o}lica de Chile,
	      Departamento de Astronom\'{i}a y Astrof\'{i}sica, Av. Vicu\~{n}a
	      Mackenna 4860, 782-0436 Macul, Santiago, Chile \\
	      }

   \date{Received xxx; accepted xxx}


   \abstract{We present high-dispersion spectra of two extremely
   massive star clusters in galactic merger remnants,
   obtained using the UVES spectrograph mounted on the ESO Very Large
   Telescope.  One cluster, W30, is located in the $\sim500$~Myr
   old merger remnant NGC~7252 and has a velocity dispersion and
   effective radius of
   $\sigma=27.5\pm2.5$~km~s$^{-1}$ and $R_{\rm eff}=9.3\pm1.7$~pc, respectively.
   The other cluster, G114, located in the $\sim3$~Gyr old merger
   remnant NGC~1316, is much more compact, $R_{\rm eff}=4.08\pm0.55$~pc, and has
   a velocity dispersion of $\sigma=42.1\pm2.8$~km~s$^{-1}$.  These measurements
   allow an estimate of the virial mass of the two clusters, yielding
   ${\cal{M}}_{dyn}(W30)=1.59(\pm0.26)\times 10^7 {\cal{M}}_{\odot}$ and
   ${\cal{M}}_{dyn}(G114)=1.64(\pm0.13)\times 10^7 {\cal{M}}_{\odot}$.  Both clusters are
   extremely massive, being more than three times heavier than the
   most massive globular clusters in the Galaxy.  For both
   clusters we measure light-to-mass ratios, which when compared to
   simple stellar population (SSP) models of the appropriate age, are
   consistent
   with a Kroupa-type stellar mass function.  Using measurements from
   the literature we find a strong age dependence on how well SSP
   models (with underlying Kroupa or Salpeter-type stellar mass
   functions) fit the light-to-mass ratio of clusters.  Based on this
   result we suggest that the large scatter in the light-to-mass ratio
   of the youngest clusters is not due to variations in the underlying
   stellar mass function, but instead to the rapidly changing
   internal dynamics 
   of young clusters.  Based on sampling statistics we argue that while
   W30 and G114 are
   extremely massive, they are consistent with being the most massive
   clusters formed in a continuous power-law cluster mass
   distribution. Finally, based on the positions of old globular
   clusters, young massive
   clusters (YMCs), ultra-compact dwarf galaxies (UCDs) and
   dwarf-globular transition objects (DGTOs) in $\kappa$-space we
   conclude that 1) 
   UCDs and DGTOs are consistent with the high mass end of star
   clusters and 2) YMCs occupy a much larger parameter space than old
   globular clusters, consistent with the idea of preferential
   disruption of star clusters.
    }      
\maketitle

%

\section{Introduction}
\label{sec:intro}

Our concept of star clusters has changed rapidly during the past
two decades.  The first resolved young clusters with masses
comparable to those of the traditional globular clusters (taken with the {\it
Hubble Space Telescope} -Holtzman et al.~1992) confirmed the
suggestions of Schweizer~(1987) that mergers of galaxies may produce
'young' globular cluster sized objects.  These results were rapidly
followed by the discovery of additional young massive clusters (YMCs) in
other galaxy mergers, as well as in dwarf, starburst, and normal
galaxies (see reviews by Whitmore~(2003) and Larsen~(2004)).  Even our
own galaxy is producing YMCs with comparable masses and sizes to those
observed in merging galaxies, e.g. Westerlund~1 (Clark et al.~2005).
The apparent ubiquity of these objects has raised the question of how
'universal' their detailed properties are, in particular concerning
their formation and subsequent evolution.

In order to address this and other questions regarding YMCs, we have
begun a programme to obtain kinematic and structural properties of
star clusters which lie at the extreme high-end of the distribution of
observed (luminous) masses.  Our first result, presented in Maraston et
al.~(2004) was for the extremely luminous star cluster, W3, in the
galactic merger remnant NGC~7252.  Combining the velocity dispersion
measured with {\it UVES} on the {\it VLT} ($45\pm5$~km~s$^{-1}$) with the
size determined from {\it HST} images (R$_{\rm eff} = 17.5\pm1.8$)
led to the dynamical mass estimate of $8(\pm2)~\times10^{7}
{\cal{M}}_{\odot}$.  This mass and that estimated from photometric
methods (see Maraston et al.~2004 for details) were in excellent
agreement, arguing that the stellar mass function within W3 was
Salpeter-like.  

Using similar techniques as were employed in the above work, some
studies have suggested that the stellar mass function in star clusters
can vary substantially (e.g. Smith \& Gallagher 2001), while others
have reported standard Kroupa~(2002) or Salpeter~(1955) type stellar mass
functions (e.g. Larsen et al.~2004).  These discrepant results have
left the question of the variance of the stellar mass function of
massive YMCs open to debate. 

Additionally, detailed knowledge of the internal dynamics and
structural parameters of YMCs has allowed a comparison between
them and other gravitationally bound systems.  In Maraston et
al.~(2004) we showed that W3 is too diffuse for its mass when compared with
old globular clusters, whereas it is too compact relative to dwarf
galaxies.  However, we also showed that W3's properties were extremely
similar to those of massive point-like objects discovered in Fornax
(Hilker et al.~1999).  Based on this similarity, Bastian et al.~(2005b)
have suggested a mechanism which may allow massive star clusters to
exist far from the main body of the host galaxy, namely the
formation of massive clusters in the tidal debris of galactic
interactions/mergers. 

Following up on these results, we have obtained high resolution {\it
UVES} optical spectra of two additional highly luminous star cluster
candidates in galactic merger remnants.  The first is the cluster W30,
the second brightest cluster in NGC~7252.  W30 has an estimated age of
$\sim300-500$~Myr and a 
metallicity between half solar and solar, estimated from
optical spectra (Schweizer \& Seitzer~1998) as well as from optical
and near-infrared photometry (Maraston et al.~2001).  The observed
magnitude of W30 is m$_{V}$(W30)$=19.46$ mag and it has a $(V-I)$ colour
of 0.63 mag
(Miller et al.~1997).  The second cluster in this study is G114 in
NGC~1316.  Based on optical and near-infrared photometry along with
near-infrared spectroscopy, Goudfrooij et al.~(2001a,b) estimate an
age of $3.0\pm0.5$~Gyr for G114.  This cluster, the brightest one
in NGC~1316, has an observed
magnitude m$_{B}$(G114)$=19.63$ mag and $(B-I)$ colour 1.87 mag.  Throughout
this work we adopt the distances to NGC~7252 and NGC~1316 which were
used in Maraston et al.~(2004) and Goudfrooij et al.~(2001a,b)
respectively, namely 64.4 and 22.9 Mpc.  Figures~\ref{w30-pa} \&
\ref{114-pa} show the {\it HST/WFPC2} planetary camera chip images of
NGC~7252 and NGC~1316, 
respectively, along with the slit sizes and positions used.  In
Fig.~\ref{w30-pa} we also mark the massive star cluster W3. 

The luminosities of G114 and W30 ( assuming a Salpeter or
Kroupa-type stellar IMF) imply that they have extremely high
masses, more
than 3 times that of the most massive globular cluster in the
Galaxy, $\omega$Cen which has a mass of $\sim3\times10^6~{\cal{M}}_{\odot}$
(Meylan \& Mayor~1986). 

In this work we investigate the structural and kinematic properties of
these two massive star clusters in order to determine how they relate
to  ``normal'' young and old globular clusters.  In \S~\ref{sizes}
we measure the effective radii of the two clusters using
high-resolution {\it HST} imaging, and in \S~\ref{sec:sigma} we
determine their velocity dispersions.  We combine these results in
\S~\ref{sec:mass} to estimate the dynamical mass of the clusters.
In \S~\ref{sec:discussion} we discuss the implications of our results
in terms of the underlying stellar mass function of the clusters, and
compare their properties with other bound stellar systems.  Finally,
we summarise the results and present our conclusions in
\S~\ref{sec:conclusions}.

\begin{figure}[!htb]
 \begin{center}
      \caption{$F555W$ image of the centre of NGC~7252 showing the slit
     size and position, as well as cluster W3 for reference.  The
     image is 740
     pixels on a side which corresponds to a linear distance of
     $\sim$10.5 kpc at the assumed distance of 64.4 Mpc.}
         \label{w30-pa}
      \end{center}
   \end{figure}

\begin{figure}[!htb]
 \begin{center}
      \caption{$F450W$ image of the centre of NGC~1316 showing the slit
     position.  The image is 800
     pixels on a side which corresponds to a linear distance of
     $\sim$4 kpc at the assumed distance of 22.9 Mpc.  North is down
     and east is to the right in this image.}
         \label{114-pa}
      \end{center}
   \end{figure}

\section{Determination of the cluster sizes}
\label{sizes}

Structural parameters for clusters G114 and W30 were measured on {\it
Wide Field Planetary Camera 2 (WFPC2)} and on {\it Advanced Camera for
Surveys (ACS) HST} images.  The images for W30 are presented in detail
in Miller et al.~(1997).  Here we note that W30 is located on the {\it
Planetary Camera} chip.  For G114 we measured the size on both the
{\it WFPC2} (data presented in Goudfrooij et al.~2001b) and {\it ACS}
(presented in Goudfrooij et al.~2004) images.

Sizes were found using the {\it ISHAPE} routine of Larsen (1999). This
routine convolves the PSF with a specified model profile of varying
sizes and fits it directly to the images.  The outputs of this routine
for the best fitting model are the FWHM of the major axis, the minor
to major axis ratio, and the goodness of fit.  For
the {\it WFPC2} images we used a PSF generated by {\it TinyTim}
(Krist \& Hook 1997) at the exact location of the cluster on the chip.  We
used the drizzled {\it ACS} images to eliminate the geometric
distortion of the camera.  The PSF for each filter of the {\it ACS} images was
constructed using sources from {\it ACS} observations of the globular cluster
47-Tuc.  

For each cluster we fit multiple profiles to the images in each of the
observed filters, including King, Moffat, and Gaussian profiles.  

\subsection{NGC~1316:G114}

Using the $F450W$ {\it WFPC2} observations, we fit eight profiles to the
image of the cluster.  The results are shown in
Table~\ref{table:size114}.  In addition to the set profiles, we also
fit a Moffat profile with a variable index.  
The best fitting model is one with an index 1.76.  Using
this profile we measure an effective radius of 4.08~pc, which is
remarkably close to the mean of the size determined using all the
other profiles (4.10~pc).  We estimate the error on the size as the
standard deviation of the size measured for the different profiles,
0.25~pc.

We have also measured the size of G114 on $F555W$ and $F814W$
{\it HST-ACS} images.  In Table~\ref{table:size114} we show the
results for the fits using the best fitting profile from the {\it WFPC2}
observations.  We find that the size determined on the {\it
F555W}$_{ACS}$ images is $\sim15$\% larger than that found on the
WFPC2 images.  However the size measured on the $F814W$$_{ACS}$
images is $\sim15$\% smaller than on the {\it WFPC2} images.  

As the three images all give approximately the same result we conclude
that G114 is resolved, and we assign the size of $R_{eff}$(G114)$ =
4.08\pm0.55$~pc.

\begin{table}[!htb]
{\scriptsize
\parbox[b]{9cm}{
\centering
\caption[]{Effective radius of NGC~1316:G114 for various models as
determined from the $F450W$ WFPC2 image.  See text for details.  The
best fitting model (lowest $\chi^2$) is shown in bold.}
\begin{tabular}{c c c c c}
\hline\hline
\noalign{\smallskip}
{\bf WFPC2} & Type & index & $R_{\rm eff}$(pc) & $\chi^2 / \chi^2_{best}$$^{b}$ \\
\hline
 & Gauss & -- & 4.34  &   1.93\\
 & King & 5 &  4.27  &    1.54\\
 & King & 15 & 3.89 &     1.13\\
 & King & 30 & 3.66   &   1.24\\
 & King & 100 & 4.14  &   1.87\\
 & Moffat & 1.5 & 4.34  &  1.09\\
 & Moffat & 2.5 & 4.09  &  1.14\\
\hline
& \\
 &  {\bf Moffat$^a$} & {\bf 1.76} & {\bf 4.08} & {\bf 1.0} \\
& \\
\hline
{\bf ACS} & Filter & Type & index & $R_{\rm eff}$  \\

& $F555W$ & Moffat & 1.76 & 4.83 $\pm0.25$ \\
& $F814W$ & Moffat & 1.76 & 3.68 $\pm0.1$ \\

\noalign{\smallskip}
\noalign{\smallskip}
\hline
\end{tabular}
\begin{list}{}{}
\item[$^{\mathrm{a}}$] Best fitting model of the WFPC2 observations.
\item[$^{\mathrm{b}}$] Calculated individually for the different cameras.
\end{list}
\label{table:size114}
}
}
\end{table}

\subsection{NGC~7252:W30}
Using the $F555W$~{\it WFPC2} image of W30, we find that the best
fitting model is a King profile, with concentration factor 100.  Using 
this profile we measure an effective radius of 8.3~pc.  However the
mean size of the other profiles gave
10.3~pc. Table~\ref{table:sizew30} gives the determined size of W30 
for different profiles.  There is a clear decreasing $\chi^2$ trend
as one goes to smaller radii (i.e. closer to the best fitting
radius). However, a King profile with concentration factor 30
along with a Moffat profile with index 1.5 are also acceptable fits in
terms of their $\chi^{2}$.
We therefore estimate the effective radius of W30 to be 9.3$\pm$1.7
pc, where the error of 1.7~pc convers the full range of acceptable
profile fits.

\begin{table}[!htb]
{\scriptsize
\parbox[b]{9cm}{
\centering
\caption[]{Effective radius of NGC~7252:W30 for various models as
determined from the $F555W$ WFPC2 image (PC chip).  See text for details.  The
best fitting model (lowest $\chi^2$) is shown in bold.}
\begin{tabular}{c c c c c}
\hline\hline
\noalign{\smallskip}
{\bf WFPC2} & Type & index & $R_{\rm eff}$(pc) & $\chi^2 / \chi^2_{best}$ \\
\hline
 & Gauss & -- & 13.2  &   3.67\\
 & King & 5 &  12.5  &    2.93\\
 & King & 15 & 11.0 &     1.82\\
 & King & 30 & 10.2   &   1.31\\
 & {\bf King} & {\bf 100} & {\bf 8.3}  &  {\bf 1.00} \\
 & Moffat & 1.5 & 11.0  &  1.21\\
 & Moffat & 2.5 & 12.0  &  2.19\\

\noalign{\smallskip}
\noalign{\smallskip}
\hline
\end{tabular}
\label{table:sizew30}
}
}
\end{table}

\section{Velocity Dispersion}
\label{sec:sigma}

We observed NGC~7252:W30 and NGC~1316:G114 with the UltraViolet
Echelle Spectrograph ({\it UVES}) mounted on the ESO/VLT on the nights of
Sept.~13-16th, 2004.  We used the red arm, CD\#3 grating centred on
5200\AA.  This resulted in a wavelength coverage from 4200\AA~ to
6200\AA~ and a resolution of $\approx 5$ km~s$^{-1}$ at
5200\AA.   The data were reduced and extracted using the on-line {\it UVES}
pipeline with the relevant bias and flat-frames.  Cosmic-rays were
also removed using the pipeline.  After extraction,
each spectrum was corrected to the helio-centric velocity frame, and
summed to create the total spectrum for each cluster. The total
exposure times were 8.67 hours on G114 and 25.1 hours on W30.
Figures~\ref{w30-pa} \&~\ref{114-pa} show the positions and lengths of
the {\it UVES} slits for clusters W30 and G114 respectively, superimposed on
{\it HST} planetary camera images.  We note that the background near
the positions of each of these clusters is devoid of spurious sources, which allowed
a clear background subtraction (determined using a spline function) at
the position of the clusters. 

Additionally, we observed several template stars which were used to
complement our existing template catalogue (Maraston et al.~2004),
namely HR~35~(F4~V), HR~8709~(A4~V), HD~203638~(K0~III), and
HD~212574~(A6~V) where the designation in the brackets refers to the
spectral type of the star.  The stars were reduced in the same way as
described above.

The determination of the velocity dispersion of each of the clusters
was carried out in exactly the same way as was done for W3, which is
described in detail in Maraston et al.~(2004).  In summary we used an
adapted version of the Fourier Correlation Quotient (FCQ, Bender~1990)
method as implemented by Bender et al.~(1994), using templates chosen to
match the stellar populations within each cluster.   The templates
were chosen to have temperatures and gravities 
(i.e. luminosity classes) appropriate to stars at the main-sequence
turn-off point and giants stars in synthetic
stellar populations of the same age as each cluster. Their
contributions to the composite template were weighted using the weights
predicted by the SSPs at the appropriate age (Maraston, 2005, e.g. her
Fig.~13).

\subsection{NGC~1316:G114}

The results of the determination of the velocity dispersion for G114
is shown in Table~\ref{table:sigma114}.  Due to the high S/N ratio of
the data and the large number of metal lines in the optical part of
the spectrum (due to the dominance of cool stars at the cluster's age
of $\sim3$~Gyr) we were able
to use the full spectral range to determine the velocity dispersion.
We note that the results do not depend crucially on the assumed stellar
template.  The adopted one-dimensional velocity dispersion for G114
is $\sigma(G114) = 42.1 \pm 2.8$~km~s$^{-1}$, which is the average over the
full wavelength range of Templates~2 and 3, which should be the
closest match to the actual stellar population based on the models of
Maraston~(2005).  Fig.~\ref{orders-114} shows the blue section of the
observed spectrum of G114
(black), the best fitting broadened stellar
template (red) and the difference between the two (green).  All
spectra shown in this work have been divided by the continuum and had
a value of one subtracted from them, to place the average value at zero.

We have also measured the heliocentric line-of-sight velocity of G114, 
$$v(G114) = 1292 \pm 3~{\rm km~s^{-1}}.$$  This is consistent with the
results of Goudfrooij et al.~(2001a), who measured $1306 \pm 26$~km~s$^{-1}$.

\begin{figure*}[!htb]
 \begin{center}
      \caption{The observed (observed divided by the continuum minus 1
     and smoothed by 15 pixels) spectrum of G114
     is plotted in 
     black.  The red shows the best fitting template, broadened by
     42.1~km~s$^{-1}$.  The green is the difference between the two, shifted
     downward by a constant for clarity.  This is just one
     region that we fitted, Table~\ref{table:sigma114} shows all the
     regions that were used. The large scale undulations
     ($\sim~30$\AA) are due to the response of the individual orders
     of the echelle spectra. Since the scale length of these
     variations is much larger than those of the spectral features of
     interest ($<$~5~\AA) they do not affect the derived velocity
     dispersion.}
         \label{orders-114}
      \end{center}
   \end{figure*}

\begin{table*}[!htb]
{\scriptsize
\parbox[b]{12cm}{
\centering
\caption[]{Velocity dispersion measurements of NGC~1316:G114.}
\begin{tabular}{c c c c c c c }
\hline\hline
\noalign{\smallskip}
Template & Spectral Type  & \multicolumn{5}{c}{$\sigma$ (km~s$^{-1}$)}  \\
         &               & 4360-5115\AA & 5245-5432\AA & 5486-5855\AA
         &5943-6113\AA & Average\\
\hline
HR~35 & F4~V & 41.8 & 47.3 & 43.9 & 43.4 & 44.1\\

HD~203638 & K0~III & 43.6 & 41.9  & 48.0 & 47.6 & 45.3\\

Template 1$^{a}$ & -- &  39.9 & 36.3 & 41.0 & 45.6 & 40.7\\

Template 2$^{b}$ & -- &   40.3 & 37.8 & 40.7 & 44.7 & 40.9\\

Template 3$^{c}$ & -- &  41.1 & 41.4 & 45.2 & 45.6 & 43.4\\

\hline

\noalign{\smallskip}
\noalign{\smallskip}
\hline
\end{tabular}
\begin{list}{}{}
\item[$^{\mathrm{a}}$] 50\% K0~III star and 50\% F4~V star
\item[$^{\mathrm{b}}$] 60\% K0~III star and 40\% F4~V star
\item[$^{\mathrm{c}}$] 70\% K0~III star and 30\% F4~V star
\end{list}
\label{table:sigma114}
}
}
\end{table*}

\subsection{NGC~7252:W30}

Based on comparisons between the optical/near-infrared colours (Miller
et al.~1997, Maraston et al.~2001) and optical spectroscopy (Schweizer
\& Seitzer~1998), W30 and W3 appear to have very similar ages and
metallicities.  Because of this, we have used the same stellar
template to determine the velocity dispersion of W30 as we used for W3
(Maraston et al.~2004).  Similarly as was done for W3, we have limited our analysis to
the region red-ward of H$\beta$, in particular concentrating on the
region around the Mg triplet ($\lambda\lambda 5167, 5172, 5183$\AA)
and Fe lines ($\lambda\lambda 5270,
5335$\AA). Table~\ref{table:sigmaw30} shows the results of the fitting
on specific regions of the spectrum. We adopt the value $\sigma(W30) =
27.5 \pm 2.5$~km~s$^{-1}$ which is the average between the value found for
fitting solely on the Mg lines and fitting on the full region of
interest (between 5150\AA~and 5350\AA).

Fig.~\ref{orders-w30} shows the
spectrum of cluster W30 in the fitting region (black), the best fitting
broadened template (red), and the difference between the two (green).

The measured line of sight velocity of W30 is $$v(W30) = 4614 \pm
1~{\rm km~s^{-1}},$$ which is in very good agreement with previous
measurements, namely
$4624 \pm 17$~km~s$^{-1}$ (Schweizer \& Seitzer~1998).

\begin{figure*}[!htb]
 \begin{center}
      \caption{The observed (smoothed by 5 pixels) spectrum of W30 is
     plotted in
     black.  The red shows the best fitting template, broadened by
     28.0 km~s$^{-1}$.  The green in the difference between the two, shifted
     downward by a constant for clarity.  The top two panels
     show the entire wavelength range which was used in the fitting,
     while the bottom figure shows a blow up of the region including
     the Mg triplet, which are the strongest lines in this region.}
         \label{orders-w30}
      \end{center}
   \end{figure*}

\begin{table}[!htb]
{\scriptsize
\parbox[b]{9cm}{
\centering
\caption[]{Derived velocity dispersion for different portions of the
spectra of W30 in NGC~7252.  All values are given in km~s$^{-1}$.}
\begin{tabular}{c c c c}
\hline\hline
\noalign{\smallskip}
Mg$^{a}$ & Fe$^{b}$ & Full$^{c}$ & Adopted \\
\hline
 & & & \\
24.7  & 30.3 & 28.0 & 27.5 $\pm$2.5 \\

\noalign{\smallskip}
\noalign{\smallskip}
\hline
\end{tabular}
\begin{list}{}{}
\item[$^{\mathrm{a}}$] Just the Mg triplet, from 5156\AA~to 5187\AA.
\item[$^{\mathrm{b}}$] Red-ward of the  Mg triplet, mainly dominated by
Fe lines, from 5218\AA~to 5351\AA.
\item[$^{\mathrm{c}}$] The full spectrum used, from 5155\AA~to 5324\AA.

\end{list}
\label{table:sigmaw30}
}
}
\end{table}

\section{Dynamical Masses}
\label{sec:mass}

Assuming that the clusters are in virial equilibrium, we can estimate
their virial masses using the relation
  \begin{equation}  {\cal{M}}_{\rm dyn} = \eta\frac{ \sigma^{2}_{\rm
x} r_{\rm eff} }{ G}  \label{eqn:mass}  \end{equation}
(Spitzer 1987, p. 11-12) where $\eta$ is a dimensionless parameter which depends on the cluster
profile adopted.  Here we adopt $\eta = 9.75$, which is approximately
valid for most of the globular clusters in the Milky Way.  However, we
will return to this adopted value, and its consequences in \S~\ref{sec:lit}.
Inserting the
numbers derived above we find that 
$$ {\cal{M}}_{\rm dyn}(G114) = 1.64 (\pm 0.13) \times 10^{7} {\cal{M}}_{\odot}$$
$$ {\cal{M}}_{\rm dyn}(W30) = 1.59 (\pm 0.26) \times 10^{7} {\cal{M}}_{\odot}$$
where the errors do not include uncertainties in the distance to the
clusters.  The adopted properties of W3, W30, and G114 are shown in
Table~\ref{table:info}. 

We note, however, that the value of the constant in
Eq.~\ref{eqn:mass} known as $\eta$, may
change as a function of age of the cluster (Boily et al.~2005).  This
effect will be most dramatic in the first ~30 Myr of a cluster's
lifetime, and the size of the effect depends on the surface mass density of
the cluster in the sense that the clusters with the highest densities
will be the most affected.  We will return to this point in
Section~\ref{sec:lit}. 

These results confirm the extremely large masses of G114 and W30.
However, it is important to put these clusters in the context of the
mass functions of the full cluster systems of their respective galaxies
in order to test to what extent they really are outliers.  \footnote{As noted
in Miller et al.~(1997) and Goudfrooij et al.~(2001a,b) both NGC~7252
and NGC~1316 have large populations of luminous star clusters which
appear to be coeval with the merger of the host galaxies.}

To do this we have performed a series of monte-carlo tests of the
mass function of cluster populations.  We assume an initial  
power-law mass distribution of the clusters within each galaxy of the form
$N({\cal{M}}){\rm d}{{\cal{M}}}\propto {\cal{M}}^{-\alpha}{\rm d}{{\cal{M}}}$, with $\alpha \sim2$
(e.g. Miller et al.~1997) and also that this distribution gets filled
randomly.  In the case of the NGC~7252 system, W3 (the most massive
cluster in the NGC~7252) is $\sim5$ times 
more massive than W30 (the second most massive cluster in this
system).  Under these conditions, we expect that $\sim20$\% of the
realizations of the
cluster populations will have a factor of five or greater between the
most massive and the second most massive clusters within the
system\footnote{We have also implicitly assumed that the ratio between
the lowest mass cluster and the highest mass cluster is $\ll0.001.$}.
The third brightest cluster in NGC~7252 system (W6) is $\sim0.2$ magnitudes
fainter than W30, corresponding to a mass difference of only $\sim17$\%
(assuming a common age).  Hence, we see that while W3 and W30 are
extremely massive clusters, they are compatible with being the most massive
clusters of a continuous power-law distribution, which we note
continues to the detection limit.

The intermediate aged cluster population of NGC~1316 can also be
readily explained by the same argument as above, as the second and
third most massive clusters in this system (G114 is the most massive)
are only 0.47 and 0.54 magnitudes fainter respectively.  This
corresponds to less than a factor of two in the luminosity (and mass
assuming that the clusters have similar ages and metallicities).  A
difference of two or greater in the ratio of the most massive and second most
massive clusters was found in $\sim50$\% of the realizations of a
cluster population.  Goudfrooij et
al.~(2004) have reported that the bright end of the luminosity
function (which we assume to represent the mass function) is well
approximated by a power-law of the type used above.

Since NGC\,7252:W3, NGC\,7252:W30, and NGC\,1316:G114 can
be readily 
understood through sampling statistics we will assume that they are
simply the most massive clusters of a continuous distribution.
In \S~\ref{sec:kappa} we will use this assumption and the detailed
properties of 
these clusters to understand more enigmatic objects, namely
the dwarf galaxy transition objects (DGTOs) and ultra-compact dwarf
galaxies (UCDs).

\begin{table*}[!htb]
\begin{centering}
{\scriptsize
\parbox[b]{18cm}{
\centering
\caption[]{The properties of the massive star clusters.}
\begin{tabular}{c c c c c c}
\hline\hline
\noalign{\smallskip}
Name  & M$_V$$^{a}$ & R$_{\rm eff}$ & $\sigma$ & age & Mass \\
      & (mag) &  (pc)         & (km~s$^{-1}$)   & Gyr &  ($10^{7}{\cal{M}}_{\odot}$) \\
\hline
NGC~7252:W3 & $-$16.2 & $17.5\pm1.8$ & $45\pm5$ & $0.4\pm0.15$ &
$8\pm2$ \\
NGC~7252:W30 & $-$14.6 & $9.3\pm1.7$ & $27.5\pm2.5$ & $0.4\pm0.15$ & $1.59\pm0.26$\\
NGC~1316:G114 & $-$13.0 & $4.08\pm0.55$ & $42.1\pm2.8$ & $3\pm1$ & $1.64\pm0.13$\\
\noalign{\smallskip}
\noalign{\smallskip}
\hline
\end{tabular}
\begin{list}{}{}
\item[$^{\mathrm{a}}$] We have corrected for foreground extinction of
NGC~7252 and NGC~1316 of $A_{V} = 0.03$  and
$A_{V} = 0.0$  respectively.

\end{list}
\label{table:info}
}
}
\end{centering}
\end{table*}

\section{Discussion}
\label{sec:discussion}
\subsection{Stellar mass functions}

A comparison between the mass of a cluster derived using photometric methods
and the mass determined through kinematic arguments, allows 
an independent test of the assumptions that went into each estimate.
The assumption that has garnered the most attention in recent years,
is that of the underlying stellar mass function (MF) of the cluster.  In
order to estimate the photometric mass of a cluster, mass-to-light
ratios from simple stellar population (SSP) models are required.
These, in turn, are heavily dependent on the assumed stellar mass
function.  Therefore, assuming that all the other assumptions are
valid (such as the state of equilibrium, correct extinction
determination, and stellar evolutionary tracks) any difference between
the mass of a cluster derived in these two ways is caused by the
difference between the input stellar mass function and that of the cluster.

Studies that have used this technique have reported strongly divergent
results.  For example, Smith \& Gallagher~(2001) and McCrady et al.~(2005) have
reported that the YMC M82F is deficient in low-mass stars, relative to
the standard Salpeter~(1955) (a single power-law mass distribution from the lower to
the upper mass limit) or Kroupa~(2002) type MFs (a single power-law above
1${\cal{M}}_{\odot}$ and significantly flatter below this limit).  However, Maraston et
al.~(2004), Larsen et al.~(2004), and Larsen \& Richtler~(2004) have
shown that clusters in a wide variety of galactic environments are
consistent with a Salpeter or Kroupa-type MF.

We therefore carry out this experiment for the two massive clusters
W30 and G114.  In Fig.~\ref{mlratio} we show the light-to-mass ratio
from the Maraston~(2005) models for solar metallicity and a Salpeter
(dashed line) \& Kroupa (solid line) stellar mass functions.
Over-plotted in red are the two clusters in the present study as well
as W3 from our previous study.  These three clusters all lie
impressively close to the value using a Kroupa MF.  Hence they are
most likely not deficient in low mass stars. 

\subsubsection{Cluster measurements from the literature}
\label{sec:lit}

In order to compare our results to other young clusters, we have taken
a sample of clusters with velocity dispersion and radii measurements
from the literature.  The clusters, their parameters, and the
corresponding references are listed in Table~\ref{table:literature}.
We have taken the fundamental parameters (age, extinction,
brightness, velocity dispersion, radius, and distance modulus)
directly from the given reference.  In some cases the $V$-band magnitude
was not given, although we note that all clusters with ages
greater than 20~Myr have observed V-band magnitudes. These older
clusters will constitute the main part of our comparison.   In those
cases where V-band magnitudes were not available we transformed the
given magnitude to 
the $V$-band using the colours in the Maraston~(2005) SSP
models at the appropriate age (which assume a Salpeter IMF).

\begin{figure*}[!htb]
\begin{centering}
     \includegraphics[width=12cm]{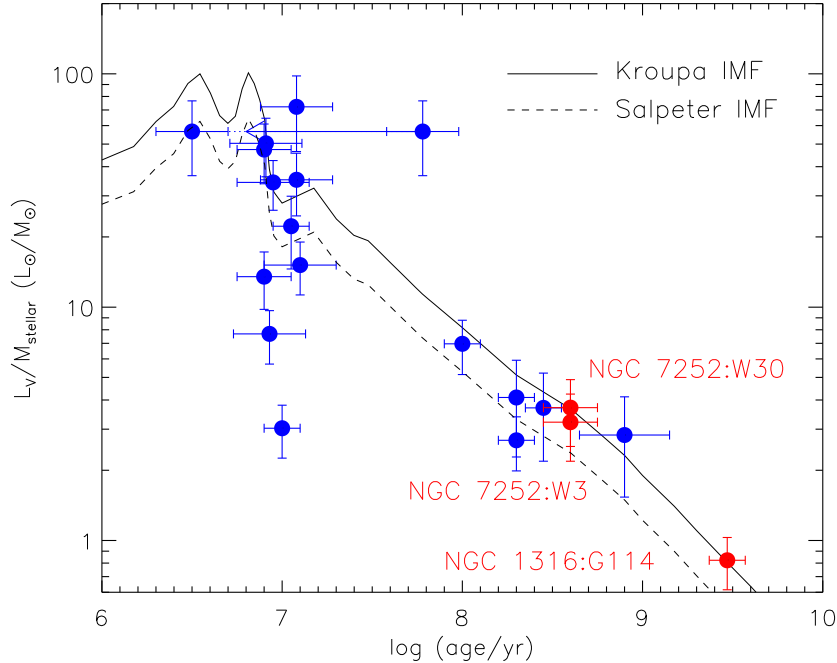}
      \caption{The derived light-to-mass ($V$-band) ratios as a
      function of the age of the clusters.  Over-plotted are the L/${\cal{M}}$
      ratios (for solar metallicity) from the Maraston (2005) SSP
      models with a Kroupa (solid line) and Salpeter (dashed line)
      stellar mass function. The clusters older than a
      few~$\times~10^7$~Myr (except M82F) are
      well fit by the models, whereas some the youngest clusters show
      significant deviation.  This may be caused by a strongly varying
      $\eta$ in the early stages of a cluster's lifetime (Boily et
      al.~2005).  For M82F, we show the position of this cluster for
      two ages, 4~Myr and 60~Myr connected with a dotted line.}
         \label{mlratio}
\end{centering}
   \end{figure*}

We have estimated the mass of each cluster using Eq.~\ref{eqn:mass}
and used their $V$-band magnitudes (and given ages) to place them in
Fig.~\ref{mlratio} (blue points).  From this figure it is clear that
the amount of deviation from standard stellar mass functions (Kroupa
or Salpeter-type) is heavily age dependent, with the older clusters
(with the exception of M82F\footnote{M82-F, located in
Fig.~\ref{mlratio} at log (age) = 7.8, may be a deviant point in the
diagram due to uncertainties in its age and extinction. Further
studies to pin down the exact values would be desirable.  Additionally,
we note that its elliptical shape, crowded environment, and peculiar
radial velocity suggest
that the cluster may have been gravitationally influenced by its
surroundings, and hence may not be in virial equilibrium.})
all consistent with a Kroupa or Salpeter-IMF and the youngest clusters
showing a large amount of scatter.  Below we outline three possible
explanations for this.

\begin{figure}[!htb]
\begin{centering}
     \includegraphics[width=9cm]{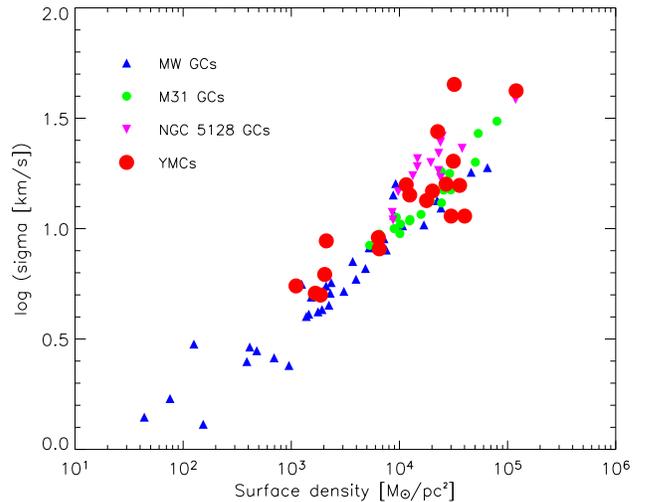}
      \caption{The measured velocity dispersion of star clusters
      versus their mean surface density within the half light radius.
      The blue triangles, small green circles, magenta upside-down
      triangles and the large red 
      circles represent old globular clusters in the Galaxy, M31, NGC
      5128, and young massive star clusters in a variety of galaxies,
      respectively.  Note that the masses have been estimated assuming
      virial equilibrium.
      However, many of the young clusters have surface
      densities greater than $10^{4}~{\cal{M}}_{\odot}/pc^2$, which is the
      regime where $\eta$ is expected to vary significantly (Boily et
      al.~2005).}
         \label{surfacedensity}
\end{centering}
   \end{figure}

A first possibility for the age-dependent scatter in
Fig.~\ref{mlratio} is that $\eta$ (the parameter
in the numerator of Eq.~\ref{eqn:mass}) changes as a
function of time (Boily et al.~2005).  This is caused by
mass-segregation in young clusters and further internal dynamical
evolution of the cluster.  The variation of $\eta$ is expected to also
be heavily dependent on the surface density of the star cluster, with
higher surface densities leading to larger variations of $\eta$.
Fig.~\ref{surfacedensity} shows the mean surface density within the
half-light radius (using the
estimated virial mass of the cluster) of star clusters vs. their
measured velocity dispersions.  The small blue filled triangles represent old globular
clusters in our galaxy (McLaughlin \& van der Marel 2005), the small
green filled circles
are globular clusters in M~31 (a collation of data from McLaughlin \&
van der Marel in prep.), the small upside-down
magenta triangles are old GCs in NGC~5128 (Martini \& Ho~2004), and the
large red circles are young massive star
clusters in a variety of galaxies (listed in 
Table~\ref{table:literature}).  All of the YMCs in
Fig.~\ref{surfacedensity} have surface densities above $10^{3}
{\cal{M}}_{\odot}$pc$^{-2}$ and most are above $10^{4}{\cal{M}}_{\odot}$pc$^{-2}$,
which is the regime where $\eta$ is expected to vary strongly (Boily
et al.~2005).

A second possibility is that many of the youngest clusters
are not in dynamical equilibrium.  This could be due to external
gravitational effects (e.g. close passages to massive GMCs).  As
clusters are born in gas-rich environments this is a likely possibility.
A lack of equilibrium could also be caused by the rapid expulsion of
the gas left over from the star formation process (assuming a
non-100\% star formation efficiency).  This can have a severe
influence on a young cluster (e.g. Boily \& Kroupa~2003).  Bastian et
al.~(2005a) have suggested that rapid gas removal may be responsible for the
dissolution of $70-90$\% of clusters within the first 10~Myr of their
lives, independent of cluster mass.  The lack of dynamical
equilibrium is also supported by the work of de Grijs, Wilkinson, \&
Tadhunter (2005) who showed that the clusters which deviate the most from
the old globular cluster M$_{V}$ vs. log$(\sigma)$ relation are found
in the highest density environments, and hence are the most likely to
be affected by interactions with the external environment.

Finally, a third possibility for the observed age-dependent scatter in
Fig.~\ref{mlratio} is that only clusters with Kroupa
or Salpeter-type stellar mass functions survive for more than
 $\sim100$~Myr. Smith \& Gallagher~(2001) suggest that if the
 cluster M82~F has a significantly top-heavy stellar IMF
 (i.e. truncated below 2-3~$M_{\odot}$) it will lack the gravitational
 potential to remain bound due to stellar evolutionary mass loss after
 2-3~Gyr.  However, in Fig.~\ref{mlratio} the clusters which deviate
 the most from Salpter or Kroupa-type stellar IMFs have light-to-mass
 ratios below the expected value.  This implies that they are
 over-abundant in low mass stars relative to Salpeter or Kroupa-type
 IMFs.  Since low mass, long-lived stars provide the gravitational
 potential to keep a cluster bound, we would expect these clusters to
 be long-lived and hence to see old clusters
 which also have light-to-mass ratios below that expected for standard
 IMFs.  Such clusters are clearly lacking in Fig.~\ref{mlratio}
 arguing against this possibility.

\subsection{More clues from $\kappa$-space}
\label{sec:kappa}

\subsubsection{The future evolution of W30 and G114}
Following the analysis by Maraston et al.~(2004) we attempt to place
W30 and G114 in the broader context of gravitationally bound stellar
systems.  For this we exploit the re-definition of the fundamental
plane known as $\kappa $-space (Bender et al. 1992) which combines the
three fundamental observable parameters (radius, surface brightness,
and velocity dispersion) into physically motivated values.  

Figure~\ref{fig:kspace} shows the position of W3 (asterisk), W30
(upward triangle), and G114 (downward triangle) in the
$\kappa1$~-~$\kappa2$ plane.  In this space $\kappa1$ traces the mass
of the system, while $\kappa2$ measures the compactness of a system
for a given mass (the product of $\cal{M}$/L and surface brightness).  For
comparison we also show the mean positions of bulges and
ellipticals~(B+E), dwarf ellipticals~(dE), M32 (all taken from Burstein et
al.~1997, and assuming H$_{0}=75$~km/sec/Mpc).  Next, we add the
point-like objects in Fornax (the ultra-compact dwarf galaxies or UCDs) (the 
average value of the four objects presented in Drinkwater et
al.~2003). We also plot the positions of {\it old} globular
clusters in the Milky Way (small blue triangles) (McLaughlin \& van der
Marel~2005), M31 (small green dots) (a collation of data from McLaughlin \&
van der Marel~in prep.), and NGC~5128
(magenta up-side triangles) (Martini \& Ho~2004), for which we assumed a
constant $(B-V)$ colour for old metal poor GCs, namely 0.7 mag. 
We also add the young clusters taken from the literature (see
Table~\ref{table:literature}), which are shown as red points.  The
large filled green squares represent the Nuclear Star Clusters (NCs)
in bulge-less disk galaxies (B\"{o}ker et al.~2004, Walcher et
al.~2005, see Table~\ref{table:literature} for details).  Finally, we
add the dwarf galaxy transition objects
(DGTOs) found in the Virgo galaxy cluster (Ha\c{s}egan et al.~2005) as
filled blue squares.  For the DGTOs we assumed that $(B-V)$=1, typical
of a 10~Gyr solar metallicity SSP.

The arrows which begin at W3, W30, and G114 represent the evolution
of the cluster in this space when the clusters are 'aged' to a common
age of 10~Gyr using the Maraston~(2005) SSP models.   Note, 
however, that the SSP model tracks do not take mass loss (and hence
fading) by evaporation or due to external perturbations into
account. 

As was found for the massive cluster NGC~7252:W3 by Maraston et
al.~(2004), we find that NGC~7252:W30 and NGC~1316:G114 will evolve
into the region of $\kappa$-space occupied by the UCDs and DGTOs.
This shows a strong similarity between the most massive star clusters and
these enigmatic objects, and may suggest that they formed through
similar mechanisms. 

Additionally, we can estimate the amount of mass loss which is
expected to occur within W30 and G114.  From the SSP models of
Maraston~(2005) we see that a cluster (assuming solar metallicity and
a Kroupa stellar IMF) is expected to lose $\sim18$\% of its mass
between the ages of 400~Myr and 10~Gyr (i.e. between the present age
of W30 and the age of globular clusters).  G114, with an age of
$\sim3$~Gyr is only expected to lose $\sim5$\% of its current mass to
stellar evolution.  Using the analytic expressions for mass loss in a
tidal field of Lamers et al.~(2005, eq.~6), we note that due to the strong
dependence on cluster mass, neither of the clusters studied here are
expected to lose a significant amount of mass due to disruption.  W30
is expected to lose $\sim~8$\% of its mass over the next 10~Gyr while
G114 is expected to lose just $\sim5$\% of its mass over the next
7~Gyr. For this calculation we have assumed t$_{4}$ (the
average time for a $10^{4} M_{\odot}$ cluster to disrupt) to be 1~Gyr,
based on the galactic GC population (Boutloukos \& Lamers~2003).
However, we note that the conclusions reached are not significantly
affected by the choice of t$_{4}$.  As these small changes
would barely be visible in Fig.~\ref{fig:kspace} and would only add
confusion, we choose not to show them.

\subsubsection{The relation between YMCs and old globular clusters}

Figure~\ref{fig:kspace-old} again shows the $\kappa1-\kappa2$
projection of $\kappa$-space, except with all of the YMCs (red points)
aged to 10~Gyr using SSP models, again assuming only passive stellar
evolution of the cluster.  Here we see that many of the YMCs have
evolved 'past' the globular cluster region and into the space occupied
by W3, W30, the UCDs and the UGTOs.  Burstein et al.~(1997) suggest
that the tightness of the GC relation in $\kappa$-space may be due to
the preferential destruction of star clusters outside a rather narrow
region of parameter space (e.g.~mass and radius, see also Fall \&~Rees~1977).  Along the same lines,
Gnedin \& Ostriker~(1997) show that only a narrow region of the
mass-radius plane of GCs is stable against disruption, and suggest
that the initial parameter distribution may have been much larger than
what is observed today.
This may be what we are seeing in Fig.~\ref{fig:kspace-old} where the
young clusters occupy a much larger region of $\kappa$-space (in terms
of mass, radius, and compactness) than their older globular cluster
counterparts. 

We note that many young star clusters are not expected to
survive for more than $\sim100$~Myr (e.g. Bastian et
al.~2005a; Fall, Chandar, \& Whitmore~2005) due to internal and external disruption mechanisms.  Thus,
it can be expected that many of the youngest YMCs in our sample will
never survive to comparable ages as the Galactic GCs.  We do not wish
to imply in Fig.~\ref{fig:kspace-old} that all the YMCs will survive
to comparable ages, only that once differences in their stellar
populations are taken into account, the YMCs occupy a much larger
parameter space than their old GC counterparts.  In particular we note
that YMCs tend to display extended envelopes in contrast to the
tidally truncated older GCs (e.g. Schweizer~2004).  The loss of such
extended envelopes (i.e. due to 
interaction with their environment) may significantly change the
position of the young clusters in Fig.~\ref{fig:kspace-old}.  As
NGC~1316:G114 is $\sim3$~Gyr old it is likely that it has already lost
its extended envelope, which may explain why it falls on the
relation for old GCs in Fig.~\ref{fig:kspace-old}.



\begin{figure*}[!htb]
 \begin{centering}
      \includegraphics[width=12cm]{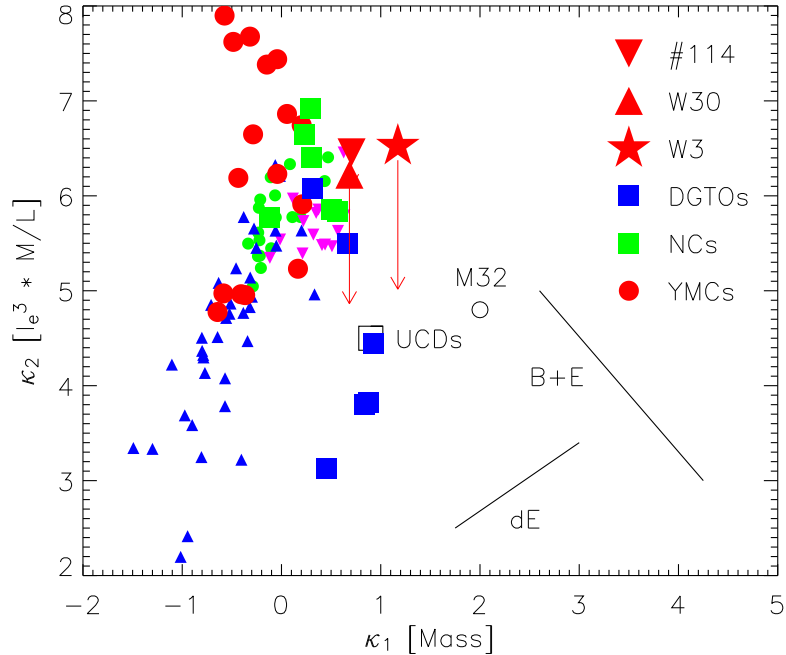}
       \caption{{\bf $\kappa$-space:}  The $\kappa$ space of stellar
 systems, in a remake of Fig.~3 of Maraston et al.~2004 (see text
 for the definitions of $\kappa1$ and $\kappa2$).  W30 and G114 are shown as red   
      triangles.   The arrows show how the 
      clusters will move in this space after ageing them to a common age
      of 10~Gyr (using SSP models). Large filled red circles
      are young clusters taken from the 
      literature (see Table~\ref{table:literature}).   Small filled blue triangles,
      green cicles, and magenta upside-down triangles are GCs from the Galaxy, M31, and
 NGC~5128 respectively (see text). The filled green squares represent
 the position of nuclear star clusters and the filled blue squares are
 the dwarf galaxy transition objects (DGTOs) found in the Virgo
 cluster (note that we assume a
 $(B-V)$=1.0 for these objects, typical of a solar metallicity 
 10~Gyr old stellar population) found in the Virgo
 cluster.  The positions of other self-gravitating objects are shown,
 ultra-compact dwarf galaxies (UCDs), M32, Bulges and Ellipticals
 (B+E), and dwarf Ellipticals (dE).}
          \label{fig:kspace}
\end{centering}
    \end{figure*}

\begin{figure*}[!htb]
 \begin{centering}
      \includegraphics[width=12cm]{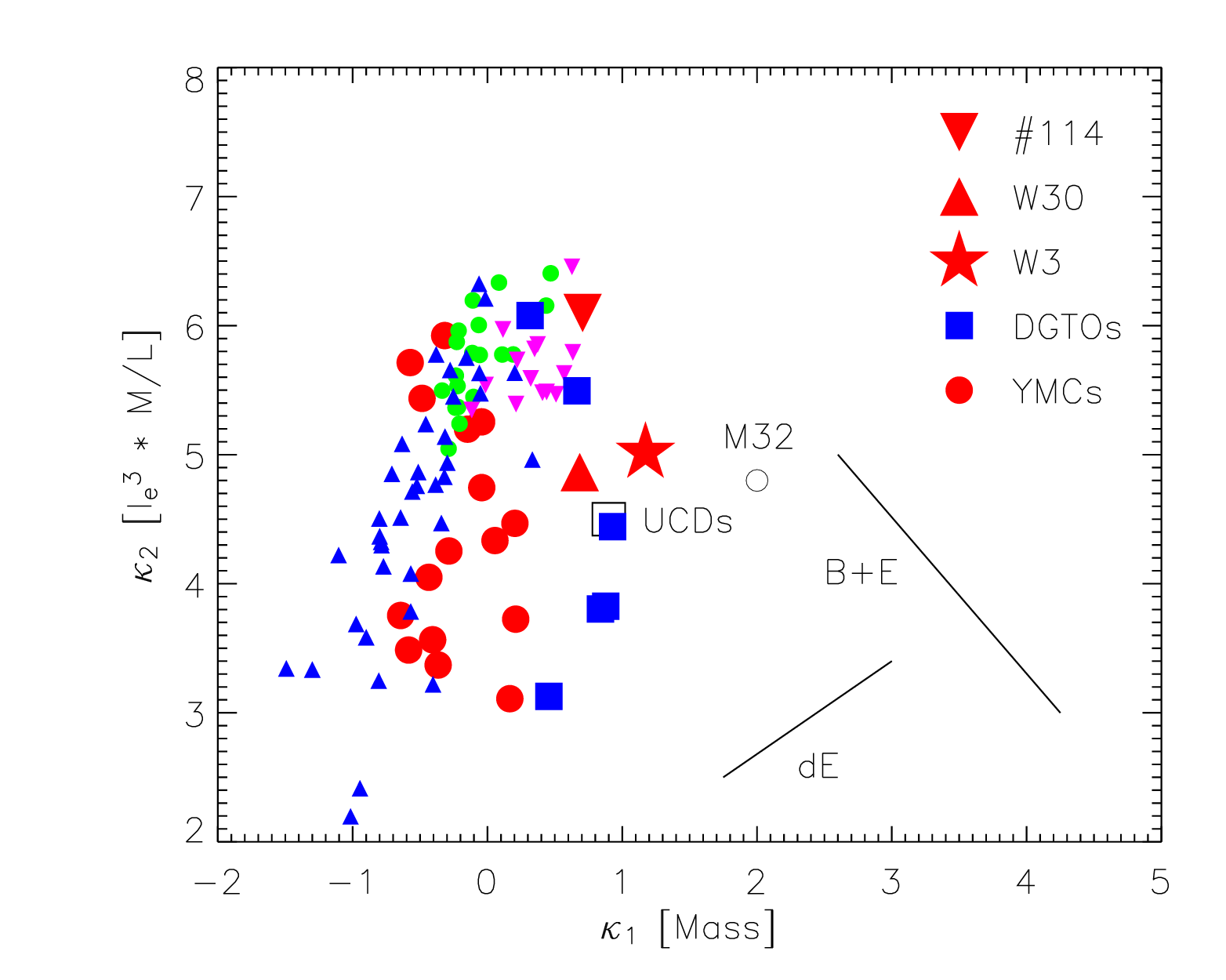}
       \caption{{\bf $\kappa$-space:} The same as
 Fig.~\ref{fig:kspace} except that all of the YMCs have been aged to
 10~Gyr using SSP models.  Again we have not included any dynamical
 evolution of the cluster (e.g. changes in radius or velocity
 dispersion).  Note that many of the YMCs will move into a similar
 region as W3, W30, the UCDs, and the DGTOs once differences in the
 stellar populations are taken into account.  The YMCs occupy a much broader
 region of $\kappa$-space than the old globular clusters.  In this
 figure we do not show the evolution of the Nuclear clusters as the
 assumption of passive stellar evolution with no additional star
 formation is clearly not valid (e.g. B\~{o}ker et al.~2004).}
          \label{fig:kspace-old}
\end{centering}
    \end{figure*}

The similarities between young massive clusters and old globular
clusters have been shown in a number of recent works
(e.g. Kissler-Patig~2004; de Grijs, Wilkinson, Tadhunter~2005).  In particular,
Kissler-Patig~(2004) showed that YMCs will follow a very similar 
$M_{V}$-$\sigma$ relation (which is one
projection of $\kappa$-space) as old GCs once the fading of the
young clusters (due to stellar evolution) is taken into account.
However, the dangers of using such a projection
can be seen when comparing Fig.~1 in Kissler-Patig~(2004) with
Figs.~\ref{fig:kspace} \&~\ref{fig:kspace-old} in the present
work.  In the $M_{V}$-$\sigma$ projection, the UCDs appear to be quite
consistent with the distribution of young and old clusters.  However,
in $\kappa$-space (which also includes information on the size) we see
that the UCDs are disjoint from the globular clusters.  Additionally,
as stated above, in $\kappa$-space the young clusters clearly occupy a
much larger parameter space than the old GCs, contrary to what is seen
in the $M_{V}$-$\sigma$ projection. Finally, we note that the
UCDs, DGTOs, and YMCs with masses above a few $\times~10^6 M_{\odot}$ all appear
very similar with respect to their scaling relations (i.e. mass,
velocity dispersion, size, and mean mass density), suggesting a common
formation mechanism (Kissler-Patig, Jord\'{a}n, Bastian~2005).

\section{Conclusions}
\label{sec:conclusions}

We have presented velocity dispersion, effective radius, and hence
dynamical mass measurements of two extremely massive clusters in
galactic merger remnants.  These results confirm that galactic mergers
can produce star clusters with masses well in excess of the most
massive globular clusters in the Milky Way.  However, we have also
shown that while these clusters are extremely massive, they are
consistent with being the high-mass end of a continuous power-law
distribution of star 
clusters, suggesting that cluster formation is (mass) scale independent.  

Comparing the light-to-mass ratios of W30 (in NGC~7252) and G114
(in NGC~1316) to those predicted by simple stellar population models
(at the ages of the clusters), shows that both of these clusters are
consistent with having Kroupa-type stellar mass functions.  Applying
the same analysis to other young clusters taken from the literature
shows that there is a significant age dependence on how well SSP
models fit the light-to-mass ratios of young clusters.  Therefore, it
is possible that the deviation from the light-to-mass ratio of young
clusters from that predicted by SSP models is not due to a varying
stellar mass function, but instead reflects the state of equilibrium
of the youngest clusters.

We have shown that W30 and G114 currently reside at the high-mass tip
of the {\it old} globular cluster distribution in $\kappa$-space (a
re-definition of the fundamental plane).  Both clusters, along with
many young clusters taken from the literature, are likely to
evolve into the space occupied by the so-called ultra-compact dwarf
galaxies (UCDs) and the dwarf galaxy transition objects (DGTOs).  This
shows that young massive clusters and UCDs/DGTOs share many similar
properties and suggests that the enigmatic UCDs/DGTOs may have formed
in a similar manner as the most massive globular clusters in mergers,
i.e., under rather violent circumstances.

Additionally, we showed that young massive star clusters will occupy a
much larger region of $\kappa$-space than presently occupied by old
globular clusters.  This is consistent with the interpretation that
star clusters are born with a larger range of parameters (e.g. radius,
mass, and compactness) than displayed by globular clusters, and
destructive processes whittle away at the initial full distribution
with only clusters which have parameters within a small range surviving
to old ages.

\begin{table*}[!p]
{\scriptsize
\parbox[b]{15cm}{
\centering
\caption[]{The properties of young massive clusters taken from the literature.}
\begin{tabular}{c c c c c c c c c c}
\hline\hline
\noalign{\smallskip}

galaxy &  ID & age &  $\Delta$age  &   M$_{\rm V}$  &  R$_{eff}$   &
$\Delta$R$_{eff}$ &  $\sigma$ &    $\Delta$$\sigma$ & reference$^{a}$    \\ 
 & & (log~yr) &  (log~yr) & (mag) & (pc) & (pc) & (km~s$^{-1}$) &
 (km~s$^{-1}$)  \\

\hline
 \\
{\bf Young Massive Star Clusters} \\
 M   82   & F 	 & 7.6(6.6)  &  0.17  &  -14.2 &    1.5   &   0.5     &  13.4   &    0.7 &   1,10 \\	
 M   82	  & MGG9 & 6.9  &  0.15  &  -13.43 &	  2.6   &   0.4	  &    15.9 &	 0.8 &   4 \\	 
 M   82	  & MGG11& 6.9  &  0.15  &  -13.23 &	  1.2   &   0.17  &    11.4 &	 0.8 &   4  \\	  	   
 NGC 1569 & A	 & 7.08 &  0.2   &  -14.1  &	  1.9	&   0.2	  &    15.7 &	 1.5 &	7,8 \\	
 NGC 1705 & 1	 & 7.08 &  0.2   &  -14.0  &	  1.6	&   0.2	  &    11.4 &	 1.5 &	1,9 \\	
 NGC 4038 & W99-15& 6.93 &  0.2  & -13.69  &	  3.6	&   0.5	  &    20.2 &	 0.7 &	6 \\	
 NGC 4038 & W99-2 & 6.95 &  0.2  & -14.79  &	  4.5	&   0.5	  &    14.2 &	 0.4 &	6 \\	
 NGC 4038 & W99-1 & 6.91 &  0.2  & -14.0   &	  3.6	&   0.5	  &    9.1  &	 0.6 &	6 \\	
 NGC 4038 & W99-16& 7.0  &  0.1  &  -12.70 &	  6.0	&   0.5	  &    15.8 &	 1.0 &	6 \\	
 NGC 4214 & 10	 & 8.3  &  0.1   &  -10.22 &	  4.33  &   0.1   &    5.1  &	 1.0 &   3 \\	 
 NGC 4214 & 13	 & 8.3  &  0.1   &  -11.68 &	  3.01  &   0.2   &    14.8 &	 1.0 &   3 \\	 
 NGC 4449 & 27	 & 8.9  &  0.25  &  -9.61  &	  3.72  &   0.32  &    5.0  &	 1.0 &   3 \\	 
 NGC 4449 & 47	 & 8.45 &  0.10  &  -10.74 &	  5.24  &   0.76  &    6.2  &	 1.0 &   3 \\	
 NGC 5236 & 502  & 8.0  &  0.1   &  -11.57 &   7.6   &   1.1	  &    5.5  &    0.2 &   2   \\       
 NGC 5236 & 805	 & 7.1  &  0.2   &  -12.17 &   2.8   &   0.4	  &    8.1  &	 0.2 &   2 \\	 
 NGC 6946 & 1447 & 7.05 &  0.10  &  -14.17 &	  10.2  &   1.6	  &    8.8  &	 1.0 &   3 \\		   
\hline
\\
NGC 1316 & G114	 & 9.47 &  0.1	 & -13.0   &	  4.1	&   0.25  &
42.1 &	 2.8 &   this work \\     
NGC 7252 & W30	 & 8.6 &  0.1	 &  -14.6  &	  9.3   &   1.7	  &
27.5 &	 2.5 &	this work \\	 
NGC 7252 & W3	 & 8.6 &  0.1   &  -16.2  &	  17.5  &   1.8	  & 45   &	 5.0 &   5 \\	
\\
 \hline
\\
{\bf Nuclear Star Clusters} & & & &  M$_{\rm I}^{b}$& \\
NGC 300 & -- & -- & -- & -11.4 & 2.9  & -- &     13.3 & 2.0 & 11,12\\
NGC 428 & -- & -- & -- & -13.1 & 3.4 & -- &	  24.4 & 3.7 & 11,12\\
NGC 1042 & -- & -- & -- & -13.1 & 1.9 & -- & 32.0 & 4.8 & 11,12\\ 
NGC 1493 & -- & -- & -- & -13.1 & 2.6 & -- & 25.0 & 3.8 & 11,12\\
NGC 3423 & -- & -- & -- & -11.8 & 4.2 & -- & 30.4 & 4.6 & 11,12\\
NGC 7793 & -- & -- & -- & -13.6 & 7.7 & -- & 24.6 & 3.7 & 11,12\\

\noalign{\smallskip}
\noalign{\smallskip}
\hline
\end{tabular}
\begin{list}{}{}
\item[$^{\mathrm{a}}$] References: 1) Smith \& Gallagher~(2001), 2)
Larsen \& Richtler~(2004), 3) Larsen et al.~(2004), 4) McCrady et
al.~(2003), 5) Maraston et al.~(2004), 6) Mengel et al.~(2002), 7) Ho
\& Filippenko~(1996b), 8) Anders et al.~(2004) 9) Ho \&
Filippenko~(1996a), 10) McCrady et al. (2005), 11) B{\" o}ker et
al. (2004), 12) Walcher et al. (2005) 
\item[$^{\mathrm{b}}$] For Fig.~\ref{fig:kspace} we have assumed $(B-I)$=1.0
which corresponds to a simple stellar population of $\sim400$~Myr and
solar metallicity, although we note that the assumed $(B-I)$
colour does not change the conclusions. 
\end{list}
\label{table:literature}
}
}
\end{table*}

\begin{acknowledgements}
We thank Marcelo Mora for his help in generating the ACS WFC PSF.
Dean McLaughlin is gratefully acknowledged for providing a uniform electronic
table of the parameters of the globular clusters in the Milky Way,
M31, and NGC 5128.  We also thank Mark Gieles for insightful
discussions, as well as Linda Smith for her help in understanding the
properties of cluster M82F.  FS gratefully acknowledges partial support from the
National Science Foundation through grant AST-0205994.
\end{acknowledgements}

\bibliographystyle{alpha}
\bibliography{../bib/astroph.bib,../bib/phd.bib,../bib/mark.bib}

\begin{thebibliography}{}

\bibitem[]{anders}  Anders, P., de Grijs, R., Fritze-v. Alvensleben,
   U., Bissantz, N.~2004, MNRAS, 347, 17
\bibitem[]{bastian} Bastian, N., Gieles, M., Lamers, H.J.G.L.M.,
   Scheepmaker, R. A., de Grijs, R.~2005a, A\&A 431, 905
\bibitem[]{bastian} Bastian, N., Hempel, M., Kissler-Patig, M.,
Homeier, N., \& Trancho, G.~2005b, A\&A, 435, 65
\bibitem[]{bender}  Bender, R.~1990, A\&A, 229, 441
\bibitem[]{bender92}  Bender, R., Burstein, D., \ Faber, S. M.~1992,
ApJ, 399, 462 
\bibitem[]{bender94}  Bender, R., Saglia, R. P., \& Gerhard, O.~1994,
MNRAS, 269, 785
\bibitem[]{boily} Boily, C.M. \& Kroupa, P.~2003, MNRAS, 338, 665
\bibitem[]{boily} Boily, C.M., Lan\c{c}on, A., Deiters, S., \& Heggie,
D.C.~2005, ApJ, 620, L27
\bibitem[]{boker} {{B{\" o}ker}, T., {Sarzi}, M., {McLaughlin}, D.~E., 
	{van der Marel}, R.~P., {Rix}, H.-W., {Ho}, L.~C., \&
	{Shields}, J.~C.} 2004, AJ, 127, 105
\bibitem{boutloukos} Boutloukos, S.G. \& Lamers, H.J.G.L.M.~2003,
	MNRAS, 338, 717
\bibitem[]{burstein} {{Burstein}, D., {Bender}, R., {Faber}, S. \&
	{Nolthenius}, R.} 1997, AJ, 114, 1365
\bibitem[]{clark} {{Clark}, J.~S., {Negueruela}, I., {Crowther},
	P.~A., \& {Goodwin}, S.~P.}, A\&A, 434, 949
\bibitem[]{degrijs} de Grijs, R., Wilkinson, M.I., \& Tadhunter,
C.N.~2005, MNRAS, 361, 311
\bibitem[]{drinkwater}  Drinkwater, M.J., Gregg, M.D., Hilker, M., et
al.~2003, Nature, 423, 519
\bibitem[]{fall77} Fall, S.M. \& Rees, M.J.~1977, MNRAS, 181, 37
\bibitem[]{fall05} Fall, S.M., Chandar R., \& Whitmore, B.C.~2005,
ApJL, in press (astro-ph/0509293)
\bibitem[]{gnedin} Gnedin, O.Y. \& Ostriker, J.P.~1997, ApJ, 474, 223
\bibitem[]{goudfrooij01a} Goudfrooij, P., Mack, J., Kissler-Patig, M.,
Meylan, G., Minniti, D.~2001a, MNRAS, 322, 643
\bibitem[]{goudfrooij01b} Goudfrooij, P., Alonso, M. V., Maraston,
C., \& Minniti, D.~2001b, MNRAS, 328, 237
\bibitem[]{goudfrooij04} Goudfrooij, P., Gilmore, D., Whitmore, B.C.,
\& Schweizer, F.~2004, ApJ, 613, 121
\bibitem[]{hasegan} {{Ha{\c s}egan}, M., {Jord{\' a}n}, A., {C{\^
o}t{\' e}}, P. et al.}~2005, ApJ, 627, 203
\bibitem[]{hilker}  Hilker, M., Infante, L., Vieira, G.,
Kissler-Patig, M., \& Richtler, T.~1999, A\&AS, 134, 75 
\bibitem[]{ho}  Ho, L.C. \& Filippenko, A. V.~1996a, ApJ, 466, L83
\bibitem[]{hob}  Ho, L.C. \& Filippenko, A. V.~1996b, ApJ, 472, 600
\bibitem[]{holtzman} Holtzman, J.A., Faber, S.M., Shaya, E.J.~et al.~1992, AJ, 103, 691
\bibitem[]{kissler-patig} Kissler-Patig, M.~2004 in The
Formation and Evolution of Massive Young Star
Clusters, ASP Conf. Ser. Vol. 322  (Astronomy Society of the Pacific -
eds. H.J.G.L.M. Lamers, L.J. Smith, A. Nota), p.~535
\bibitem[]{kissler-patig06} Kissler-Patig, M., Jord\'{a}n, A., \&
Bastian, N.~2005, A\&A submitted (Oct.)
  \bibitem[Krist \& Hook (1993)]{krist} Krist, J., \& Hook, R.~1997,
   ``The Tiny Tim User's Guide'', STScI
\bibitem[]{kroupa} Kroupa, P.~2002, Science, 295, 82
\bibitem[]{lamers} Lamers, H.J.G.L.M., Gieles, M., Bastian, N.,
Baumgardt, H., Kharchenko, N., \& Portegies Zwart, S.~2005, A\&A, in press
\bibitem[]{larsen99} Larsen, S.S.~1999, A\&AS, 139, 393
\bibitem[]{larsen04} Larsen, S.S.~2004 in Planets to Cosmology:
Essential Science in Hubble's Final Years, (ed. M. Livio) astro-ph/0408201
\bibitem[]{larsen04b} Larsen, S.S., Brodie, J.P., \& Hunter, D.A.~2004,
AJ, 128, 2295
\bibitem[]{larsen04c} Larsen, S.S. \& Richtler, T.~2004, A\&A, 427, 495
\bibitem[]{maraston} Maraston, C.~2005, MNRAS, 362, 799
\bibitem[]{maraston01}  Maraston, C., Kissler-Patig, M., Brodie, J.,
et al.~2001, A\&A, 370, 176
\bibitem[]{maraston04}  Maraston C., Bastian N., Saglia R. P.,
Kissler-Patig M., Schweizer F., \& Goudfrooij P.~2004, A\&A, 416, 467
\bibitem[]{martini} Martini, P. \&  Ho, L.C.~2004, ApJ, 610, 233
\bibitem[]{mclaughlin05} McLaughlin, D. E, \& van der Marel, R. P. 2005, ApJS, submitted
\bibitem[]{mcrady03} {{McCrady}, N., {Gilbert}, A.~M. \& {Graham},
J.~R.} 2003, ApJ, 596, 240
\bibitem[]{mccrady05} McCrady, N., Graham, J.R., \& Vacca, W.D.~2005,
ApJ, 621, 278
\bibitem[]{mengel}  Mengel, S., Lehnert, M. D., Thatte, N., \& Genzel,
R.~2002, ApJ, 550, 280
\bibitem[]{meylan} Meylan, G., \& Mayor, M.~1986, A\&A, 166, 122
\bibitem[]{miller} Miller, B.W., Whitmore, B.C., Schweizer, F., \&
Fall, S.M.~1997, AJ, 114, 2381
\bibitem[]{salpeter} Salpeter, E.E.~ApJ, 121, 161
\bibitem[]{schweizer87} Schweizer, F.~1987, in Nearly normal galaxies:
From the Planck time to the present; Proceedings of the Eighth Santa
Cruz Summer Workshop in Astronomy and Astrophysics, New York,
Springer-Verlag, p.~18
\bibitem[]{schweizer04} Schweizer, F.~2004, in The
Formation and Evolution of Massive Young Star
Clusters, ASP Conf. Ser. Vol. 322  (Astronomy Society of the Pacific -
eds. H.J.G.L.M. Lamers, L.J. Smith, A. Nota), p.~111
\bibitem[]{schweizer98}  Schweizer, F., \& Seitzer, P.~1998, AJ, 116, 2206
\bibitem[]{smith}  Smith, L.J., \& Gallagher, J.S.~2001, MNRAS, 326, 1027
\bibitem[]{spitzer} Spitzer, L.~1987, Dynamical Evolution of Globular
Clusters, Princeton Series in Astrophysics, Princeton University Press
\bibitem[]{walcher} {{Walcher}, C.~J., {van der Marel}, R.~P., {McLaughlin}, D., 
	{Rix}, H.-W., {B{\" o}ker}, T., {H{\" a}ring}, N., 
	{Ho}, L.~C., {Sarzi}, M. \& {Shields}, J.~C.}, 2005, ApJ, 618, 237
\bibitem[]{whitmore00} Whitmore, B.C.~2003, in A Decade of Hubble Space
	Telescope Science, STScI Symposium Series
14  (Cambridge University Press - eds. M. Livio, K. Noll,
\& M. Stiavelli), p.~153  

\end{thebibliography}

\end{document}